\documentstyle[aps,preprint,eqsecnum,tighten,floats,epsf,prd]{revtex}

\newcommand{\beq}{\begin{equation}}
\newcommand{\eeq}{\end{equation}}
\newcommand{\beqa}{\begin{eqnarray}}
\newcommand{\eeqa}{\end{eqnarray}}

\begin{document}

\preprint{\vbox{
\noindent  \hfill BU-HEPP-99-02\\}}
\draft

\title{Finite Volume Effects in Self Coupled Geometries}

\author{Walter Wilcox\footnote{
E-mail:~wilcoxw@baylor.edu ~$\bullet$~ Tel:
254~710~2510 ~$\bullet$~ Fax: 254~710~3878 \hfill
\null\quad WWW:~http://www.baylor.edu}}

\address{Department of Physics, Baylor
University, Waco, TX 76798-7316}

\maketitle

\begin{abstract} 
By integrating the pressure equation at the surface of a self coupled
curvilinear boundary, one may obtain asymptotic estimates of
energy shifts, which is especially useful in 
lattice QCD studies of nonrelativistic
bound states. Energy shift expressions
are found for periodic (antiperiodic) boundary conditions
on antipodal points, which require Neumann (Dirichlet) boundary 
conditions for even parity states
and Dirichlet (Neumann) boundary conditions for odd parity 
states. It is found that averaging over periodic and antiperiodic 
boundary conditions is an effective way of removing the asymptotic energy 
shifts from the boundary. Asymptotic
energy shifts from boxes with self coupled
walls are also considered and shown to be effectively
antipodal. The energy shift equations are
illustrated by the solution of the bounded harmonic oscillator
and hydrogen atoms.

\end{abstract}
\noindent

\vfill
\newpage

\section{Introduction}
\label{sec:intro}

In investigating the effects of self coupled boundaries in 
Monte Carlo lattice QCD simulations there
is a need for guidance as to the expected size and direction
of boundary-induced energy
shifts as well as strategies to
remove these effects. Starting with parameterizations of (quenched)
lattice potentials within the context of a nonrelativistic
two-body Schr\"odinger-Pauli Hamiltonian with relativistic corrections, a
recent approach uses these potentials to produce genuine
predictions of charmonium and bottomonium energy
levels\cite{boyle1}. These predictions are obtained from 
numerical solutions, using a purely spatial grid, of the 
wavefunctions plus perturbation theory. 
Using this approach, one may begin to understand finite
size effects numerically by changing the size of the
enclosing box, which is relatively easy to do, unlike the
Monte Carlo simulations\cite{boyle2}. 
Although the general case must still
be studied numerically, this article shows that the asymptotic energy
shifts from distant boundaries can be determined analytically
in terms of the unbounded wavefunctions. Where these wavefunctions are 
unavailable, it provides functional forms to compare with in potential 
type simulations, and ultimately, directly to the lattice data
once box sizes can confidently be said to be in the
asymptotic region.

The systems studied here will be described by the Schr\"odinger
equation in the interior of a volume, the surface of which
is determined by a single parameter of a general 
separable curvilinear coordinate system.
It will be assumed that the fields on the boundaries are self-coupled
in the sense of being there being either periodic or antiperiodic
boundary conditions relating the antipodal points at ${\bf r}$ and
$-{\bf r}$. We will see that periodic (antiperiodic) boundary conditions 
require Neumann (Dirichlet) boundary conditions for even parity states
and Dirichlet (Neumann) boundary conditions for odd parity 
states for parity invariant potentials. These boundary conditions 
are in general not the usual ones
applied in cubic geometries, which couple opposite walls
rather than antipodal points. However, we will see that the 
asymptotic shifts when opposite walls are coupled are
effectively antipodal. Thus the description here 
should apply to the situation of
Refs.\cite{boyle1,boyle2}, which studies confining solutions
(linear plus Coulomb) in a periodic box.

We will proceed to a derivation of the energy shift formulas
in Sections \ref{sec:momentum} and \ref{sec:parameter}.
The basic energy shift equations for Dirichlet and Neumann
boundary conditions will be derived in Section \ref{sec:basic}. 
An asymptotic form of the wavefunction
will be presented in Section \ref{sec:alternate}, which will then allow
us to integrate the differential forms, resulting in simple equations
for the energies themselves. It will be seen that the asymptotic energy
shifts for Dirichlet and Neumann cases are equal in magnitude
but opposite in direction. In Section \ref{sec:sur} we will 
make some simplifying observations regarding surface 
integrals in the energy shift formulas and 
in Section \ref{sec:box} we will explicitly consider the
usual lattice situation of cubic geometry with self coupled
walls. As examples of the use of 
these formulas, in Section \ref{sec:ex} we will study a one dimensional
example, the harmonic oscillator, and a three dimensional example,
the hydrogen atom. We will close with some general observations regarding
qualitative behaviors of energy shifts in the general case
as well as some summary comments.

\section{Scr\"odinger Momentum Tensor Derivation}
\label{sec:momentum}

One way to relate boundary induced energy shifts to properties of the 
Schr\"odinger wavefunctions is to use the momentum tensor. The Lagrangian
desity for the Schr\"odinger case may be written
\begin{equation}
{\cal L} = \frac{1}{2m}\partial_{i}\psi^{*}\partial_{i}\psi + 
(V-E)\psi^{*}\psi . \label{lagrangian}
\end{equation}
The energy, $E$, can be viewed as a Lagrange multiplier
associated with the normalization constraint, $\int d^{3}x\, \psi^{*}\psi=1$.
The Schr\"odinger equation is recovered
from the action defined by Eq.(\ref{lagrangian}) by 
variation of $\psi^{*}$. (In the time dependent case one 
concludes $\psi^{*}$ and $\psi$ 
are related by complex conjugation only after independent variations
on $\psi^{*}$ and $\psi$.) The mass parameter $m$ should be understood
to represent the reduced mass in applications to two body systems. 

The momentum tensor for complex fields may be written as

\begin{equation}
T_{ij}=\frac{\delta{\cal L}}{\delta (\partial_{i}\psi^{*})}\partial_{j}\psi^{*} 
+\frac{\delta{\cal L}}{\delta (\partial_{j}\psi^{*})}\partial_{i}\psi^{*}
-\delta_{ij}{\cal L},
\end{equation}
which is explicitly symmetric in $i$ and $j$. Using Eq.(\ref{lagrangian}),
this gives

\begin{equation}
T_{ij}=\frac{1}{2m}[\partial_{i}\psi \partial_{j}\psi^{*}+
\partial_{j}\psi\partial_{i}\psi^{*}]
-\frac{\delta_{ij}}{2m}\partial_{k}(\psi^{*}\partial_{k}\psi ).
\end{equation}
As a check, we notice that

\begin{equation}
\partial_{i}T_{ij}= -\psi^{*}\partial_{j}V\psi ,
\end{equation}
which is just a form of Ehrenfest's theorem when integrated over the
volume. The continuity equation

\begin{equation}
\frac{\partial g_{i}}{\partial t} + \partial_{j}T_{ji}=0,
\end{equation}
is satisfied in the time dependent case, where the 
momentum density is given by

\begin{equation}
g_{i}= -\frac{i}{2}(\psi^{*}\partial_{i}\psi-\psi\partial_{i}\psi^{*}),
\end{equation}
assuring momentum conservation. We will calculate energy
shifts by evaluating the pressure exerted by the wavefunction on the surface.
For this purpose consider the momentum tensor with locally normal indices:

\begin{equation}
T_{nn}=\frac{1}{2m}\partial_{n}\psi\partial_{n}\psi^{*}
-\frac{1}{2m}{\bf \partial}_{T}\psi{\bf \partial}_{T}\psi^{*}
-\frac{1}{2m}\psi^{*}{\bf \nabla}^{2}\psi .
\end{equation}
Here 
\begin{equation}
{\bf n}\cdot {\bf \nabla}\psi \equiv \partial_{n}\psi = 
\frac{1}{h_{n}}\frac{\partial\psi}{\partial n},
\end{equation}
is a locally normal derivative (${\bf n}$ is outwardly directed),
$h_{n}$ is a possible scale factor, and
${\bf \partial}_{T}$ are the tranverse derivatives. 
In the context of integrations over a closed surface,

\begin{equation}
\frac{1}{2m}\psi^{*}{\bf \nabla}^{2}\psi +
\frac{1}{2m}{\bf \partial}_{T}\psi{\bf \partial}_{T}\psi^{*} \\
\longrightarrow \frac{1}{2m}\psi^{*}\partial_{n}^{2}\psi ,
\end{equation}
giving

\begin{equation}
T_{nn}=\frac{1}{2m}|\partial_{n}\psi |^{2}
-\frac{1}{2m}\psi^{*}\partial_{n}^{2}\psi.
\label{giving}
\end{equation}

Antipodal boundary conditions are given by

\begin{eqnarray}
\psi ({\bf r})|_{s} = \pm \psi (-{\bf r})|_{s}, \label{bfirst}\\
\partial_{n}\psi ({\bf r})|_{s} = \mp \partial_{n}\psi (-{\bf r})|_{s}.
\label{bsecond}
\end{eqnarray}
where the top signs for the periodic case and the bottom signs
give the antiperiodic case. (One plugs in $-{\bf r}$ after 
the normal derivative is taken in (\ref{bsecond}).) For a parity
invariant potential, one has

\begin{equation}
\psi |_{s}=0,
\label{d}
\end{equation}
by continuity of the wavefunction at the surface for odd parity states,
and

\begin{equation}
\partial_{n}\psi |_{s}=0,
\label{n}
\end{equation}
by continuity of the first normal derivative of the wavefunction for
even parity states in the periodic case.
For antiperiodic boundary conditions, the conditions
(\ref{d}) and (\ref{n}) apply instead
to even and odd parity states, respectively. When the states can
not be classified by parity, as when an otherwise good parity system
is not centered at the origin, Eqs.(\ref{bfirst}) and (\ref{bsecond}) can no
longer be simplified.

For the Dirichlet condition Eq.(\ref{d}), one then has 

\begin{equation}
T_{nn}=\frac{1}{2m}| \partial_{n}\psi  |^{2},
\end{equation}
and

\begin{equation}
T_{nn}= -\frac{1}{2m}\psi^{*}\partial_{n}^{2}\psi ,
\end{equation}
for Neumann conditions, Eq.(\ref{n}).

Finally, we relate the energy shift to an integration over
$T_{nn}$ via
\begin{equation}
\Delta E  \equiv  \beta (a_{0}) = 
\int PdV = \int_{\infty}^{a_{0}}da \int ds \, h_{n}\, P(a),
\end{equation}
where
\begin{equation}
P(a)  =  -T_{nn}(a).
\end{equation}
Thus, in a differential sense one has

\begin{equation}
\frac{d\beta}{da}=-\int ds\, h_{n} \, T_{nn}(a),
\label{resultingin}
\end{equation}
resulting in

\begin{equation}
\frac{d\beta}{da}=-\frac{1}{2m}\int ds\, h_{n} \, | \partial_{n}\psi |^{2},
\label{dirichlet}
\end{equation}
in the Dirichlet case and

\begin{equation}
\frac{d\beta}{da}= \frac{1}{2m}\int ds\, h_{n} \,
 \psi^{*}\partial^{2}_{n}\psi ,
\label{neumann}
\end{equation}
in the Neumann. Again, in the general case without good parity,
the form from Eq.(\ref{giving}) can not be simplified.

These differential energy shift formulas are actually exact. 
The wavefunction, $\psi$, is a shorthand
for $\psi (\beta (a))|_{s}$, denoting the exact
wavefunction on the boundary determined at finite $\beta$, 
which is itself determined by the boundary parameter, $a$, 
for a given surface. As they stand, these
equations are not yet useful because we have no information on the
values of the wavefunction and it's normal derivatives on
$S$. A hint on how to proceed is provided by an alternate 
derivation of these results, which will now be presented.

\section{Parameter Variation Derivation}
\label{sec:parameter}

Let us consider the variation of the Schr\"odinger equation,

\begin{equation}
-\frac{1}{2m}{\bf \nabla}^{2}\psi + V\psi=E\psi ,
\label{schro}
\end{equation}
with respect to the parameter ({\it not} the coordinate), $a$, 
which determines the self coupling surface, 
assuming the normalization condition,
$\int d^{3}x\, \psi^{*}\psi =1$, is satisfied. (Usually the
following procedure
determines the normalization integral given $E(a)$,
but here we use it in reverse.) We have

\begin{equation}
-\frac{1}{2m}\psi^{*}{\bf \nabla}^{2}\frac{\partial\psi}{\partial a} + \psi^{*}
V\frac{\partial\psi}{\partial a} =\frac{d\beta}{da}\psi^{*}\psi+E\psi^{*}
\frac{\partial\psi}{\partial a} ,
\end{equation}
after variation followed by $\psi^{*}$ multiplication. On the other hand,
complex conjugation of Eq.(\ref{schro}) followed by  
multiplication by $\frac{\partial\psi}{\partial a}$ results in

\begin{equation}
-\frac{1}{2m}\frac{\partial\psi}{\partial a}{\bf \nabla}^{2}\psi^{*} +
\frac{\partial\psi}{\partial a} V\psi^{*}=
E\frac{\partial\psi}{\partial a}\psi^{*} .
\end{equation}
The difference produces a perfect differential form, which gives

\begin{equation}
\frac{d\beta}{da}=\frac{1}{2m}\int ds\, \left(\frac{\partial\psi}{\partial a}
\partial_{n}\psi^{*}
 - \psi^{*}\partial_{n}\frac{\partial\psi}{\partial a} \right),
\label{otherform}
\end{equation}
where $\partial_{n}$ are again local normal derivatives. (The order
of the partial derivatives in the last term is immaterial since
$a$ appears only in $\psi$.) We then
have

\begin{equation}
\frac{d\beta}{da} = \frac{1}{2m}\int ds\, \frac{\partial\psi}{\partial a}
\partial_{n} \psi^{*},
\label{altd}
\end{equation}
for Dirichlet boundary conditions, and

\begin{equation}
\frac{d\beta}{da} = -\frac{1}{2m}\int ds\,  \psi^{*}
\frac{\partial}{\partial a}\partial_{n} \psi ,
\label{altn}
\end{equation}
for the Neumann case. These results may be reconciled with 
Eqs.(\ref{dirichlet}) 
and (\ref{neumann}) by the use of expansions in the normal 
curvilinear coordinate, $n$, about the boundary.
In the Dirichlet case since the wavefunction vanishes on the surface one
must have

\begin{equation}
\psi  \approx (n-a)\frac{\partial\psi}{\partial n}|_{s},
\end{equation}
resulting in

\begin{equation}
\frac{\partial\psi}{\partial a}|_{s} = - h_{n}\partial_{n}\psi|_{s}.
\label{result1}
\end{equation}
In the Neumann case since the first normal derivative vanishes one has

\begin{equation}
\partial_{n}\psi\approx
(n-a)\frac{\partial}{\partial n}\partial_{n}\psi|_{s},
\end{equation}
giving

\begin{equation}
\frac{\partial}{\partial a}\partial_{n}\psi|_{s}=
- h_{n}\partial^{2}_{n}\psi|_{s}.
\label{awk}
\end{equation}
This reconciles Eqs.(\ref{altd}) and 
(\ref{altn}) with 
(\ref{dirichlet}) and (\ref{neumann}). In the general case with no
good parity, one must proceed differently. One may expand
about the boundary the quantity,

\begin{eqnarray}
\left( \frac{\partial_{n}\psi ({\bf r})}{\psi^{*} ({\bf r})} + 
\frac{\partial_{n}\psi ({-\bf r})}{\psi^{*} (-{\bf r})}
\right),\nonumber
\end{eqnarray}
(for $|\psi|_{s}$ nonvanishing) which also 
vanishes on S. Then one learns 
that Eq.(\ref{resultingin}), with the general $T_{nn}$ from 
(\ref{giving}), and
Eq.(\ref{otherform}) are equivalent when the 
antipode terms are grouped together
inside the surface integral. One also learns that one may replace

\begin{equation}
\partial_{n}^{2}\psi |_{s}
 \rightarrow \frac{1}{h_{n}^{2}}
\frac{\partial^{2}\psi}{\partial {n}^{2}}|_{s},
\end{equation}
anywhere in these expressions.

\section{Basic Formulas}
\label{sec:basic}
The use of the $a$ parameter allows one to derive 
a differential equation
for the energy shift, $\beta$, and simplifies the Neumann expression.
We begin with a change of variables from $a$ to $\beta (a)$:

\begin{equation}
\frac{\partial\psi}{\partial a}=\frac{d\beta(a)}{da}
\frac{\partial\psi}{\partial\beta} .
\label{first}
\end{equation}
In the Dirichlet case one may expand the boundary condition, which determines
the energy values, in $\beta$,

\begin{equation}
0=\psi |_{s} \approx \psi (0)|_{s} +\beta
\frac{\partial\psi}{\partial\beta} |_{s},
\end{equation}
($\psi (0)$ specifies the unbounded wavefunction)
allowing the approximate replacement

\begin{equation}
\frac{\partial\psi}{\partial\beta} |_{s}\approx -\frac{1}{\beta}\psi (0) |_{s}.
\label{third}
\end{equation}
When (\ref{result1}), (\ref{first}) and (\ref{third}) are are used in 
(\ref{altd}), the variables $\beta$ and $a$ separate and we have

\begin{equation}
\int^{\beta(a_{0})}\frac{d\beta}{\beta^{2}}
\approx -2m \int^{a_{0}}\frac{da}{\int ds\, h_{n}^{-1}|\psi(0)|^{2}}.
\end{equation}
This identifies the anayltic form for $1/\beta (a_{0})$ up to a constant.
Thus, asymptotically one has

\begin{equation}
\beta^{D}(a_{0})\approx \left( 2m\int^{a_{0}}\frac{da}
{\int ds\, h_{n}^{-1}|\psi(0)|^{2}}
\right)^{-1} \label{Dshift}.
\end{equation}

The Neumann case proceeds similarly, beginning with

\begin{equation}
\frac{\partial}{\partial a}\partial_{n}\psi = 
\frac{d\beta (a)}{da}\frac{\partial}{\partial\beta}
\partial_{n}\psi.
\label{begin}
\end{equation}
The boundary condition is again expanded in $\beta$,

\begin{equation}
0=\partial_{n} \psi |_{s}\approx \partial_{n}\psi(0) |_{s} 
+\beta\frac{\partial}{\partial\beta}\partial_{n}\psi |_{s},
\end{equation}
resulting in

\begin{equation}
\frac{\partial}{\partial\beta}\partial_{n}\psi |_{s}\approx 
-\frac{1}{\beta}\partial_{n}\psi(0) |_{s}.
\label{result2}
\end{equation}
As we will see in the next Section under very general conditions
for good parity states,

\begin{equation}
\psi (\beta(a))|_{s} \approx 2\psi(0)|_{s}.
\label{general}
\end{equation}
When (\ref{begin}), (\ref{result2}) and (\ref{general}) are 
used in (\ref{altn}), no integration is needed 
and we simply obtain

\begin{equation}
\beta^{N} (a_{0})\approx \frac{1}{m}\int ds_{0}\, 
\psi^{*}(0)\partial_{n}\psi (0),
\label{1}
\end{equation}
for the energy shift. Note that we have not
assummed that the wavefunctions 
are separable. For a general confining ellipsoidal surface
it can be shown that in the asymptotic limit the appropriate
scale factor goes to one\cite{morse}. In the following
we will also see that it is always possible to choose surfaces
for which these factors are effectively unity. 
Therefore we will set $h_{n}=1$ in the following.

\section{Alternate Approach}
\label{sec:alternate}

An alternate approach determines the energy shifts from
the form of the asypmtotic wavefunction. Assuming that the potential, $V$,
is spherically symmetric far from the region where the
unbounded wavefunction exists (measured by it's average radius, $<r>$,
say), we have the approximate wave equation

\begin{equation}
\left[ \frac{d^{2}}{dr^{2}} - \kappa^{2}(r) \right]\psi (\beta (a)) = 0,
\label{waveE}
\end{equation}
which is solved by

\begin{eqnarray}
\psi(\beta (a))  =  \psi (0)|_{s} 
(e^{-\int_{r_{a}}^{r}\kappa(r')dr'}
\pm e^{\int_{r_{a}}^{r}\kappa(r')dr'}),
\end{eqnarray}
where $r_{a} = r|_{s}$ and where the upper sign goes with
Neumann boundary conditions and the lower goes with Dirichlet and the
second term gives a vanishingly small contribution far from the boundary.
In the immediate vincinity of the boundary, the unbounded wavefunction
may be characterized as

\begin{equation}
\psi(0)\equiv\psi (0)|_{s} e^{-\int_{r_{a}}^{r}\kappa(r')dr'}.
\label{asymp}
\end{equation}
Thus we conclude in the Dirichlet case,

\begin{equation}
\partial_{n}\psi ( \beta (a))|_{s} \approx 2\partial_{n}\psi(0)|_{s},
\end{equation}
giving the alternate expression,

\begin{equation}
\beta^{D} (a_{0})\approx -\frac{2}{m}\int^{a_{0}}_{\infty} da 
\int ds\, \left| 
\partial_{n}\psi \right|^{2}.
\label{altD}
\end{equation}

Similarly, in the Neumann case one has (\ref{general}) above as well as

\begin{equation}
\partial^{2}_{n}\psi (\beta (a))|_{s}\approx
2\partial^{2}_{n}\psi(0)|_{s},
\end{equation}
giving the alternate form

\begin{equation}
\beta^{N} (a_{0})\approx  
\frac{2}{m}\int^{a_{0}}_{\infty} da  \int ds\, \psi^{*} (0)
\partial^{2}_{n}\psi (0).
\label{altN}
\end{equation}
The next Section will show that use of the aymptotic forms of 
the wavefunctions allows the Dirichlet
or Neumann energy shift formulas to be integrated
and results in$^{1}$

\begin{equation}
\beta^{D}(a_{0})\approx -\beta^{N}(a_{0}) \approx -\frac{1}{m}\int ds_{0}\,
\psi^{*}(0)\partial_{n}\psi (0).
\label{both}
\end{equation}
This equation can also be interpreted as the energy shift in systems in higher
or lower dimensions as long as the basic asymptotic wave equation
(\ref{waveE}) holds in the appropriate variable.

\section{Surface Integrals}
\label{sec:sur}

In the last two Sections we have derived differently appearing formulas
for the energy shifts. These will be reconciled in this Section. In the
process we will learn about the surface integrals in these formulas
in the general situation when the enclosing surface is not spherical.

The reconcilation of (\ref{1}) and (\ref{altN}) 
in the Neumann case is accomplished by use of the asymptotic
solutions to Eq.(\ref{waveE}). The condition $r_{a}\gg\,\, <r>$
in each quantum state (or the appropriate statement in one dimension)
and the existence of the asymptotic wave equation, Eq.(\ref{waveE}),
insures that we can ignore any $r$ dependence in the 
unbounded wavefunction, $\psi (0)$, when integrating or
taking normal partial derivatives,
other than from the exponential in Eq.(\ref{asymp}). Writing the
general asymptotic form of the unbounded wavefunction 
(still not assuming separability; $\Omega$ represents the
angular variables) as

\begin{equation}
\psi (0) = N(r,\Omega)e^{-\int_{0}^{r}dr' \, \kappa (r')},
\label{gen}
\end{equation}
one can show that either (\ref{1}) or (\ref{altN}) give the common form,

\begin{equation}
\beta^{N}(a_{0}) \approx -\frac{1}{m}\int ds_{0} \,  \,
\kappa (r_{a})\, \partial_{n}r \,|\psi (0)|^{2},
\label{common}
\end{equation}
This form may be further reduced because of the
presence of the exponential in the surface integral.
This will suppress contributions from regions of the surface
not in the immediate vicinity of the points of
closest approach to the potential center. Let us call $r_{\perp}$ 
this closest distance. Thus, we may pull the
$r$ dependent quantities
outside of the spatial integral, frozen at their values
for $r=r_{\perp}$,

\begin{equation}
\beta^{N}(a_{0})\approx -\frac{1}{m} \,
\kappa (r_{\perp})\, \frac{\partial r_{\perp}}{\partial n} \int ds_{0} \,
|\psi (0)|^{2}.
\label{frozen}
\end{equation}

In the Dirichlet case,
Eq.(\ref{altD}) may similarly be reduced to the negative of
the right hand side of
Eq.(\ref{frozen}) using the asymptotic form, Eq.(\ref{gen}).
On the other hand, starting from (\ref{Dshift}) one may write

\begin{equation}
\int ds \, |\psi (0)|^{2} = e^{-2\int_{0}^{r_{\perp}}
\kappa (r')dr'}\int ds \, 
|N(r_{a},\Omega)|^{2}\,e^{-2\int_{r_{\perp}}^{r_{a}}\kappa (r')dr'}.
\label{inter}
\end{equation}
The surface integral in (\ref{inter}) can be treated as a constant in 
the boundary integral in (\ref{Dshift}) (verified below), 
again resulting in the negative of
the right hand side of Eq.(\ref{frozen}) for $\beta^{D}(a_{0})$.

Eq.(\ref{frozen}) is still unnecessarily complicated. Since
the near points on the boundary will dominate the surface integral,
there are three cases that can occur for the locus of these
points: sphere, circle and point. In the sphere case
one simply has

\begin{equation}
\beta^{N}_{sphere}(a_{0}) \approx -\frac{1}{m} \,
\kappa (r_{\perp})\,  \int ds_{0} \, |\psi (0)|^{2}.
\label{sphere}
\end{equation}
In the case where the near points are a circle, one 
can construct a local cylindrical surface for the
integration, giving

\begin{eqnarray}
\int ds \,  |N(r_{a},\Omega)|^{2}\,
e^{-2\int_{r_{\perp}}^{r_{a}}\kappa (r')dr'} & \approx &
r_{\perp} \int _{0}^{2\pi}d\phi \int_{-\infty}^{\infty} dz \, 
|N(r_{\perp},\Omega_{\perp})|^{2}\,e^{-2\kappa_{\perp}
(r_{a}-r_{\perp})},  \nonumber \\
& \approx &
r_{\perp}\int_{0}^{2\pi} d\phi\, |N(r_{\perp},\Omega_{\perp})|^{2}\,
\int_{-\infty}^{\infty} dz \, 
e^{-\frac{\kappa_{\perp}}{r_{\perp}}z^{2}}, \nonumber \\
& = &
\sqrt{\frac{\pi r_{\perp}^{3}}{\kappa (r_{\perp})}}
\int_{0}^{2\pi} d\phi\, |N(r_{\perp},\Omega_{\perp})|^{2}.
\end{eqnarray}
This results in

\begin{equation}
\beta^{N}_{circle}(a_{0}) \approx -\frac{1}{m} \,
\sqrt{\pi r_{\perp}^{3}\kappa (r_{\perp})}
\int_{0}^{2\pi} d\phi \, |\psi_{\perp} (0)|^{2},
\label{circle}
\end{equation}
where $\psi_{\perp} (0)$ is the
wavefunction along the $r_{\perp}$ points. 
Likewise, if the nearest equidistant points are discrete,
one can consider a locally flat surface and evaluate

\begin{eqnarray}
\int ds \,  |N(r_{a},\Omega)|^{2}\,
e^{-2\int_{r_{\perp}}^{r_{a}}\kappa (r')dr'} & \approx &
2\pi \sum_{i} \, |N_{i}(r_{\perp},\Omega_{\perp})|^{2} 
\int _{0}^{\infty }d\rho\rho \,e^{-2\kappa_{\perp}
(r_{a}-r_{\perp})},  \nonumber \\
& \approx &
2\pi \sum_{i} \, |N_{i}(r_{\perp},\Omega_{\perp})|^{2}
\int _{0}^{\infty }d\rho\rho \,e^{-\frac{\kappa_{\perp}}{r_{\perp}}
\rho^{2}}, \nonumber \\
& = &
\frac{\pi r_{\perp}}{\kappa (r_{\perp})}\sum_{i} \,
 |N_{i}(r_{\perp},\Omega_{\perp})|^{2} ,
\end{eqnarray}
where the sum is over the equidistant points. This results in

\begin{equation}
\beta^{N}_{points}(a_{0}) \approx
-\frac{\pi r_{\perp}}{m}
\sum_{i} \, |\psi_{i\perp} (0)|^{2}.
\label{points}
\end{equation}
This last case is the appropriate one
for a cubic box enclosure with antipodal boundary conditions.

\section{Boxes with Self Coupled Sides}
\label{sec:box}

The antipodal boundary conditions considered
here are not the ones used in Monte Carlo lattice simulations
since they distinguish one spatial point as the \lq\lq center\rq\rq . These
simulations generally use periodic or antiperiodic couplings on
opposite box sides. Nevertheless, the antipodal conditions
are {\it effectively} implemented in such simulations because of
the local nature of the surface integrals. Let us
see how this comes about.

In general one should consider both continuity of $\psi$ and
$\partial_{n}\psi$ on the surfaces
in constructing the
energy shift formulas for boxes with self coupled sides. Let
us consider the energy shift from the $x=\pm a$
surfaces of the box for a parity invariant,
centered potential. The asymptotic wavefunction
with antipodal boundary conditions 
in this case can be written

\begin{equation}
\psi(\beta (a)) \approx  \psi (r_{a},+)\, 
e^{-\int_{r_{a}}^{r}\kappa(r')dr'}
\pm \psi (r_{a},-)\, e^{\int_{r_{a}}^{r}\kappa(r')dr'},
\label{boxpsi}
\end{equation}
where the upper sign is for periodic and the lower is for 
the antiperiodic case, and where

\begin{equation} 
\psi (r_{a},+) \equiv N(r_{a},\Omega )\,
e^{-\int_{0}^{r_{a}}\kappa (r')dr'},
\end{equation}
gives the general form of the unbounded wavefunction with $r$
replaced by $r_{a}=\sqrt{a^{2}+y^{2}+z^{2}}$.
$\psi (r_{a},-)$ is the same as above
except with $\Omega$ reflected through the $x=0$ plane.
The second term on the right of Eq.(\ref{boxpsi})
will be exponentially suppressed in the interior of the box. 
Now using the general results, Eqs.(\ref{giving}) and 
(\ref{resultingin}), and integrating over $a$ 
using the asymptotic form (\ref{boxpsi}), we have

\begin{equation}
\beta^{box}(a_{0})\approx \mp \frac{1}{m} \,
 \int ds_{0} \, \kappa (r_{a})\, \partial_{n} r\, 
{\rm Re}(\psi^{*} (r_{a},+)\psi (r_{a},-)).
\label{persides}
\end{equation}
($S_{0}$ indicates both sides of the box.)
We will get the full energy shift by integrating over the other
sides of the box in a similar manner. 
Because of the local nature of these integrations,
the contributing portions of the box sides are almost antipodal,
and the factors $\kappa (r_{a})$ and $\partial_{n} r$ 
can come outside of the surface integral.
Then, for wavefunctions with good parity,

\begin{equation}
\psi_{\perp} (r_{a},+)= \pm \psi_{\perp} (r_{a},-),
\end{equation}
Eq.(\ref{persides}) in the periodic case
just gives the results of the last Section for
Neumann and Dirichlet boundary conditions, respectively. 
Antiperidicity just reverses these results. Near the box
edges the representation (\ref{boxpsi}) fails to be correct, but the
local nature of the surface integrations allows it
as an approximation. When the potential is not centered or the
states do not have good parity, similar considerations
show that periodic and antiperiodic energy shifts
are negative and equal, at least in the case where
one point is much closer to the force center than the other.

\section{Examples}
\label{sec:ex}

We will consider two examples of the use of the energy
shift formulas, the one dimensional harmonic oscillator and
the hydrogen atom.

For the harmonic oscillator, the asymptotic form of
the wavefunctions are given by (${\bar x}\equiv 
(m\omega)^{-1/2}$; ${\tilde x} \equiv x_{0}/{\bar x}$;
$x_{0}$ is the boundary distance
from the origin))

\begin{equation}
\psi_{N}({\tilde x}) \equiv \frac{{\tilde x}^{N}
e^{-\frac{1}{2}{\tilde x}^{2}}}
{\sqrt{{\bar x}N!(2)^{-N}\sqrt{\pi}}}.
\label{wave}
\end{equation}
Use of (\ref{wave}) in Eq.(\ref{both}) 
(the \lq\lq surface\rq\rq\, integral simply supplies a factor of two)
gives immediately that 

\begin{equation}
\beta^{D}_{N}(x_{0}) \approx \frac{(2)^{N+1}
{\tilde x}^{2N+1}e^{-{\tilde x}^{2}}}{m{\bar x}^{2}N!\sqrt{\pi}}.
\label{betaD1}
\end{equation}

On the other hand, the explicit energy shifts for 
the harmonic oscillator may be found from the
general forms of the solutions, proportional to confluent 
hypergeometic functions, using the boundary condition on the wavefunction. 
The form of the wavefunction is\cite{cush}

\begin{equation}
\psi^{even} ({\tilde x}) \sim e^{-\frac{1}{2}
{\tilde x}^{2}}F(\frac{-\lambda+1}{4}\,|\frac{1}{2}|\,{\tilde x}^{2}),
\end{equation}
for the even parity states, and the energies are 
$E=\lambda/2 m{\bar x}^{2}$. Expanding $\lambda$ as

\begin{equation}
\lambda = (4N+1) + 2m{\bar x}^{2}\Delta E,
\end{equation}
($N=0,1,2,\dots$) one has in the Dirichlet case , $\psi^{even} ({\tilde x})
=0$, that

\begin{equation}
\Delta E^{D}_{N}(x_{0}) \approx  -\frac{F(-N\,|\frac{1}{2}|\,
{\tilde x}^{2})}
{2m{\bar x}^{2}\frac{\partial}{\partial\lambda} F(-N-\frac{\lambda}{4}\, |
\frac{1}{2}|\, {\tilde x}^{2})|_{\lambda=0}}.
\end{equation}
Asymptotically,

\begin{equation}
F(-N\, |\frac{1}{2}|\, {\tilde x}^{2}) \longrightarrow 
\frac{(-1)^{N}}{\left( \frac{1}{2} \right)_{N}}
{\tilde x}^{2},
\end{equation}
where
\begin{equation}
(a)_{N} \equiv \prod_{i=0}^{N-1} (a+i) = \frac{\Gamma (N+a)}
{\Gamma (a)}.
\label{gammas}
\end{equation}
The method of taking derivatives of confluent hypergeometric functions
in Ref.\cite{wilcox} is used,
neglecting polynomials in $x_{0}$ up to order $2N$
in $\frac{\partial}{\partial\lambda} F(-N-\frac{\lambda}{4},
\frac{1}{2},\frac{1}{2}{\tilde x}^{2})|_{\lambda=0}$.
After a shift in the summand and considerable reduction, 
this can be put into the form

\begin{equation}
\Delta E^{D}_{N}(x_{0}) \approx 
\frac{2{\tilde x}}{N!}\left[\sum_{\tau=2N+1}^{\infty}
\frac{{\tilde x}^{2\tau -4N+1}}
{(N+\frac{1}{2})_{\tau-2N}(\tau-2N)_{N+1}}
\right]^{-1}.
\label{infinite}
\end{equation}
In dropping the polynomial contribution, we must have that 
the major contribution to the infinite sum in (\ref{infinite})
occurs for $\tau \gg 2N$. Since 
$<N|x^{2}|N>={\bar x}^{2}(N+\frac{1}{2})$ in the unbounded state N and
because the major contribution
in the infinite sum actually occurs for $\tau \approx {\tilde x}^{2}$,
we have that $x_{0}^{2}\gg\, 2<N|x^{2}|N>$ for these asymptotic forms to hold, 
as would be expected. 

One notices that the Eq.(\ref{betaD1}) is odd in $x_{0}$, while 
(\ref{infinite}) is even, so a term by term 
comparison of the two expressions (expanding 
$e^{-{\tilde x}^{2}}= (e^{{\tilde x}^{2}})^{-1}$ in a power series)  
is not possible. However, the asymptotic limit of the two
sums are identical. This can be established by using

\begin{eqnarray}
(N+\frac{1}{2})_{\tau-2N}(\tau  - 2N)_{N+1} &\approx &
\frac{(\frac{1}{2})_{\tau+1}2^{N}}{(2N-1)!!},\label{approx}\\
\sum_{\tau=2N+1}^{\infty}
\frac{{\tilde x}^{2\tau+1}}{(\frac{1}{2})_{\tau+1}}& \longrightarrow &
-\sqrt{\pi}\,e^{{\tilde x}^{2}}.
\label{limit}
\end{eqnarray}
Eq.(\ref{approx}) holds for $\tau \gg 2N$ and
Eq.(\ref{limit}) can be justified by using Eq.(\ref{gammas}),
replacing the sum on $\tau$ with an
integral, and doing the shift $\tau\rightarrow \tau-\frac{1}{2}$
(neglecting a polynomial of order $4N+1$ in ${\tilde x}$). Finally, 
using (\ref{approx}) and 
(\ref{limit}) in (\ref{betaD1}), we now have that
\begin{equation}
\Delta  E^{D}_{N}(x_{0})= \beta^{D}_{2N}(x_{0}),
\end{equation}
as is appropriate for the even parity states. Going back to the 
general form of the wavefunction, one may also
verify that asymptotically $\Delta E^{N}_{N}(x_{0})=-\Delta 
E^{D}_{N}(x_{0})$ from the 
Neumann boundary condition, $\partial_{x}\psi^{even} (x)|_{x=x_{0}}=0$.

For our second example we consider a hydrogen atom bounded by a
sphere. The asymptotic form of the 
radial wavefunctions for the hydrogen atom
are (${\tilde r} \equiv r_{0}/Na_{0}$; \lq\lq $a_{0}$\rq\rq\
is the Bohr radius)

\begin{equation}
R_{NL}({\tilde r}) \approx \left( \frac{2}{Na_{0}} \right)^{3/2}
\frac{({2\tilde r})^{N-1} e^{-{\tilde r}}}{\sqrt{2N(N+L)!(N-L-1)!}}.
\end{equation}
This immediately implies the Dirichlet energy shifts

\begin{equation}
\beta^{D}_{NL}(r_{0}) \approx \frac{1}{ma_{0}^{2}N^{3}}\frac{
({2\tilde r})^{2N}
 e^{-2{\tilde r}}}{(N+L)!(N-L-1)!},
\end{equation}
from Eq.(\ref{both}). Obviously, the surface integral was trivial
in this case.

The explicit form of the
energy shifts for the hydrogen atom have 
previously been evaluated in Ref.\cite{wilcox}
in the Dirichlet case. This was done by expanding the general form 
of the radial wavefunction

\begin{equation}
R(r) \sim e^{-\frac{r}{na_{0}}}r^{L}F(L-n+1\,|2L+2|\,
\frac{2r}{na_{0}}),
\end{equation}
with energy

\begin{equation}
E= -(2ma_{0}^{2}n^{2})^{-1},
\end{equation}
to first order in a Taylor series in the energy shift,
 
\begin{equation}
n\approx N+ma_{0}^{2}N^{3}\Delta E,
\end{equation}
($N=1,2,3,\dots$) and then solving for $\Delta E^{D}_{NL}$ 
with boundary condition
$R(r_{0})=0$. One obtains after reductions,

\begin{equation}
\Delta E^{D}_{NL}(r_{0}) \approx \left[ ma_{0}^{2}N^{3}(N+L)!(N-L-1)!
\sum_{\tau=2N+1}^{\infty} \frac{(\tau -2N-1)!(2{\tilde r})^{\tau-2N}}
{(\tau-L-N-1)!(\tau +L-N)!}
\right]^{-1},
\end{equation}
using the asymptotic form of the wavefunctions and
neglecting a finite polynomial series of order $N-L-1$ in $r_{0}$
in the denominator. For $\tau\gg N$ or $L$ (guaranteed for ${\tilde r}
\gg N^{2}$), one has

\begin{equation}
\frac{(\tau -2N-1)!}
{(\tau-L-N-1)!(\tau +L-N)!} \longrightarrow \frac{1}{\tau !},
\end{equation}
which results in

\begin{equation}
\Delta E^{D}_{NL}(r_{0})=\beta^{D}_{NL}(r_{0}),
\end{equation}
when another polynomial of order $2N$ in $r_{0}$ is neglected in
forming the exponential, $e^{2{\tilde r}}$. 
Once again, one may verify that the Neumann shift is just the negative of the 
Dirichlet one.

\section{Comments and Summary}
\label{sec:com}

A number of comments may be made about energy shifts in general
from the above formulas.

$\bullet$ Averaging over Dirichlet and Neumann boundary conditions
by averaging over periodic and antiperiodic boundary conditions
in lattice simulations, is an effective way of removing the 
asymptotic energy shifts for nonrelativistic bound systems with
good parity. This point can
and should be tested numerically in potential simulations. 
It is the assumption of good parity which allows this
statement to be made, and is not true for
antipodal boundary conditions in general.

$\bullet$ The exact formula, Eq.(\ref{dirichlet}), shows that the
energy shift is never negative in the case of Dirichlet
boundary conditions, which corresponds to the usual meaning
of a bounded system. This makes sense from the point of view of
the the extra kinetic energy generated
from the Heisenberg uncertainty principle.

$\bullet$ The exact differential energy shift in the case of 
Neumann boundary conditions for a central potential bounded by
a sphere can be written as

\begin{equation}
\frac{d\beta}{da}= \int ds\, \psi^{*}(\beta (a))\left[
V(a)-E(a)+\frac{L(L+1)}{2ma^{2}}
\right]\psi (\beta (a)).
\label{Nshifts}
\end{equation}
Assuming the right hand side is dominated by the (positive) spherical
confinement potential, $V(a)$, at large $a$, one immediately
sees that the asymptotic energy shifts are negative as $a$ is
decreased. For purely Coulombic systems
the right hand side is instead dominated by $E(a)$, 
which is negative, leading again to asymptotic
negative energy shifts. At extremely small $a$, one expects
the positive Neumann box energies, $E\sim 1/a^{2}$ to
dominate, leading to positive energy shifts. Thus, in the simplest
scenario one expects a minimum in the energy at some intermediate
value of $a$. This seems to be what is seen in the
numerical solutions corresponding to Neumann boundary
conditions (periodic even parity states) in Ref.\cite{boyle2},
although the confining geometry is a cube, not a sphere.
Actual proof of this scenario in spherical and other geometries however is 
difficult and more complicated types of behaviors can not be ruled out.

As illustrations of the derived formulas we considered two examples
of centered potentials: a one dimensional harmonic oscillator 
and a spherically bounded 
three dimensional hydrogen atom. The asymptotic energy shifts
were calculated by the use of the derived formulas and verified
from the known forms of the wavefuctions for the even parity
oscillator and general quantum number Coulomb systems. 
Although we have not illustrated the more general case
of a confining nonspherical boundary, 
formulas have been given for the asymptotic surface integrals 
encountered. The common situation of a confining
box with self coupled sides is also amenable to numerical 
calculation using a formula of Section \ref{sec:box}. 
Thus, the formalism here may readily be applied
to lattice simulations of nonrelativistic bound systems in the usual
box geometry. When applied
to truly periodic systems, not self coupled as in lattice studies,
these results actually estimate the location of one edge of the
Bloch energy bands which emerge\cite{hasen}.

\section*{Acknowledgments}
This work originated in a conversation with P. Boyle regarding
the results in Ref.\cite{boyle2}.
The author gratefully acknowledges support from
NSF Grant No.\ PHY-9722073, the Department of 
Physics and Astronomy, University of Kentucky, as well as from
the Special Research Centre for the Subatomic Structure of Matter,
University of Adelaide.

\newpage
\center{
{\large FOOTNOTE}}
\vspace{.5cm}
\begin{enumerate}

\item An alternate unintegrated expression, 
coming from the 
Dirichlet expression, Eq.(\ref{Dshift}), is

\begin{eqnarray}
\beta^{D} (a_{0}) \approx \frac{1}{2m}\int ds\, \left[ \int^{a_{0}} 
\frac{da}{|\psi (0)|^{2}} \right]^{-1}. \nonumber
\end{eqnarray}
\end{enumerate}

\end{document}